\documentclass[aps,pre,onecolumn,amsmath,amssymb,nofootinbib,superscriptaddress,floatfix,showpacs]{revtex4} %
\usepackage[pdftex,colorlinks,citecolor=blue,linkcolor=blue,bookmarks=false,urlcolor=blue]{hyperref} %
\usepackage{amsfonts}
\usepackage{amsmath}
\usepackage{graphicx}
\usepackage{dcolumn}
\usepackage{bm}

\newcommand{\N}{\text{N}}

\newcommand{\cG}{ {\cal G} }
\newcommand{\Sec}[1]{Sec.~\ref{#1}}

\begin{document}

\title{Gene-Mating Dynamic Evolution Theory II: \\ 
Global stability of N-gender-mating polyploid systems}
\author{ Juven C. Wang }
\email{juven@ias.edu}
 
\affiliation{Department of Physics, Massachusetts Institute of Technology, Cambridge, MA 02139, USA}
\affiliation{Perimeter Institute for Theoretical Physics, Waterloo, ON, N2L 2Y5, Canada}
\affiliation{School of Natural Sciences, Institute for Advanced Study, Einstein Drive, Princeton, NJ 08540, USA }
\affiliation{Center of Mathematical Sciences and Applications, Harvard University, MA 02138, USA}
\affiliation{Department of Physics, National Taiwan University, 
Taipei 10617, Taiwan}



\begin{abstract}
Extending the previous 2-gender dioecious diploid gene-mating evolution model, 
we attempt to answer ``whether the Hardy-Weinberg global stability and the exact analytic dynamical solutions can be found 
in the generalized N-gender N-polyploid gene-mating system with an arbitrary number of alleles?''
For a 2-gender gene-mating evolution model, a pair of male and female determines the trait of their
offspring. Each of the pair contributes one inherited character, the allele, 
to combine into the genotype of their offspring. Hence, for an N-gender N-polypoid gene-mating
 model, each of N different genders contributes one
allele to combine into the genotype of their offspring.
We exactly solve the analytic solution of N-gender-mating $(n+1)$-alleles governing highly-nonlinear coupled differential equations in the genotype frequency parameter space for any 
positive integer N and $n$. 
For an analogy, the 2-gender to N-gender gene-mating equation generalization is analogs to the 2-body collision to the N-body collision
Boltzmann equations with continuous distribution functions of
\emph{discretized} variables instead of \emph{continuous} variables.
We find their globally stable solution as a continuous manifold and find no chaos. 
Our solution implies that the Laws of Nature, under our assumptions, provide
no obstruction and no chaos to support an N-gender gene-mating stable system.


\end{abstract}



\maketitle

\tableofcontents






\section{\label{sec:level1}Introduction}

Dynamical (time-dependent) evolution of population percentages labeled by biological traits  
or genetic characteristics 
in a population
system is a primary focus of studies for population genetics and evolutionary
biology (\cite{{Wright},{Fisher},{Crowbook},book1, book2,book3} and references therein). 
Their governing laws involve genetics, which can be traced back to Mendel's foundational work in 1865 \cite{Mendel}.

In a previous work \cite{1410.3456}, we had surveyed the time-evolution of genotype frequency 
in a  2-gender dioecious diploid population system. 
We derive our governing equations
based on our four fundamental assumptions,

\noindent (a)The mating of gene-holders is (approximately) closed in a
population system.

\noindent (b)The mating probability for certain gene-type to another
gene-type, due to average-and-random mating mechanism, is proportional to
the product of their population (e.g., quadratic forms for 2-gender, cubic forms for 3-gender, and their generalizations: degree-N polynomial forms for N-gender mating.). 

\noindent (c)The probability of a gene-type for newborn generation obeys the
Mendelian inheritance.

\noindent (d)The accumulation of human population obeys the exponential
growth law, for both the birth and the death evolution.

We had derived the governing equations in the parameter space of the genotype frequencies, 
and we had solved the analytical solution
of the governing equations within a single locus and arbitrary number of alleles mating process. 
However, our previous work only focus
on 2-gender mating, normally one male and one female. 
We can analytically prove the global stability and Hardy-Weinberg equilibrium \cite{{Hardy},{Weinberg}}.
For biological and mathematical interests, we ponder whether the N-gender case,
where N is an arbitrary integer number, will bring interesting and surprising new results. In
other words, we like to consider a mating system of N different
genders. The mating process of the system will be successful if and only if
there is a group of N different genders participate in the mating process.
For a certain genotype (or a biological trait), 
each of these different N genders contributes an allele,
together combine into the inherited genotype of their
offspring. 

The questions we ask and aim to address are: Whether 2-gender mating is special
for N=2, so that it assures the global stability for the system under time evolution?
Whether 2-gender mating is special so that there exist exact analytic
solutions? Whether there is chaos for N-gender mating system at some integer
N? Is it possible to solve the analytic solution for arbitrary N? 
Is there a specific integer number M, such that for gender number smaller than M, the system has
the global stability; 
if the gender number is larger than M, 
then the chaos appears (from global stability to chaos)? If so, what is the integer M?

Since the 1-gender mating is just a trivial asexual reproduction, and the 2-gender
mating is a case with a global stable solution, one might expect that
the 2-gender mating is a transitional case for the dynamical properties of N-gender mating system. 
In other words, for N $>2$, there might exist chaos or more complicated time-evolution phase diagrams.
However, as we will show later, for the general N-gender gene-mating
system, there is always a global stable manifold as long as the system satisfies the previous four assumptions. 
Throughout the article, we focuses on the standard equations without
considering the mutation and natural selection.
From the mathematic viewpoint,
the highly symmetric governing equations result in a continuous global
stable manifold. This higher dimensional curved manifold is a global stable
attractor, under time evolution attracting all points in the Euclidean fiber attached at every
manifold point. The stable manifold and the fibers together form a fiber bundle
completely filling in the whole genotype-frequency parameter space. We
solve the analytical solutions exactly for the general N-gender ($n+1$)-alleles
gene-mating system (See  \cite{1410.3456} and the footnote \ref{footnote:alleles}).

One essential remark is that the coefficients of the governing equations for N-gender mating can be derived explicitly from the coefficient of ${\N \choose k_0, k_1, \ldots, k_n}$ 
in the well-known multinomial theorem:
\begin{equation} \label{eq:multinomial}
(\cG_0 + \cG_1 + \cdots + \cG_n)^\N = \underset{k_0, k_1, \ldots, k_n\ \in \mathbb{N}}{\sum_{k_0 + k_1 + \cdots +k_n=\N}} {\N \choose k_0, k_1, \ldots, k_n}.
  \cG_0^{k_0}  \cG_1^{k_1} \cdots \cG_n^{k_n}. 
\end{equation}
Here we denote the inherited character (commonly known as the allele) as $\cG_j$ where $j=0,1,\dots,n$ for $(n+1)$ types of alleles.\footnote{For example, in  \cite{1410.3456}, the ($n+1$)-alleles can be regarded as
$n$ dominant alleles and 1 recessive allele in a single locus.
We can denote the 1 recessive allele as $\cG_0$, and 
the $n$ dominant alleles as $\cG_1,  \cdots , \cG_n$.
\label{footnote:alleles}} 
We can count the number of population carrying a certain
genotype $\cG_0^{k_0}  \cG_1^{k_1}   \cdots \cG_n^{k_n}$ (from the set of alleles $\cG_j$ inherited from each of the N-gender parents)
as $ G_{
  \underset{k_{0}}{\underbrace{\alpha _{0}\cdots \alpha _{0}}} 
 \underset{k_{1}}{\underbrace{\alpha _{1}\cdots \alpha _{1}}}
 \dots
\dots
\underset{
k_{m}}{\underbrace{\alpha _{m}\cdots \alpha _{m}}}}$, where $\alpha_j$ is an 
labeling inherited with this specific allele $\cG_j$.  
So we have a map from the set of alleles to a certain
genotype, then to a total number of population with this genotype:
\begin{equation}
\cG_0^{k_0}  \cG_1^{k_1}   \cdots \cG_n^{k_n} \to
 G_{
  \underset{k_{0}}{\underbrace{\alpha _{0}\cdots \alpha _{0}}} %
 \underset{k_{1}}{\underbrace{\alpha _{1}\cdots \alpha _{1}}}
\dots
\dots
\underset{%
k_{m}}{\underbrace{\alpha _{m}\cdots \alpha _{m}}}}.
\end{equation} 
So far we only define our notations, later in main sections, we will write down the governing dynamical equations for the population evolution as in \cite{1410.3456}.
We will also renormalize the ${\small G}_{\dots }$ via dividing it by the total population ${P}$,
\begin{eqnarray} 
\sum {\small G_{
  \underset{k_{0}}{\underbrace{\alpha _{0}\cdots \alpha _{0}}} %
 \underset{k_{1}}{\underbrace{\alpha _{1}\cdots \alpha _{1}}}
\dots
\dots
\underset{%
k_{m}}{\underbrace{\alpha _{m}\cdots \alpha _{m}}}}}
&\equiv& P ,\\
\left(\frac{{\small G_{
  \underset{k_{0}}{\underbrace{\alpha _{0}\cdots \alpha _{0}}} %
 \underset{k_{1}}{\underbrace{\alpha _{1}\cdots \alpha _{1}}}
\dots
\dots
\underset{%
k_{m}}{\underbrace{\alpha _{m}\cdots \alpha _{m}}}}}}{P}
\right)&\equiv&{\small G'_{
  \underset{k_{0}}{\underbrace{\alpha _{0}\cdots \alpha _{0}}} %
 \underset{k_{1}}{\underbrace{\alpha _{1}\cdots \alpha _{1}}}
\dots
\dots
\underset{%
k_{m}}{\underbrace{\alpha _{m}\cdots \alpha _{m}}}}} \to {\small G_{
  \underset{k_{0}}{\underbrace{\alpha _{0}\cdots \alpha _{0}}} %
 \underset{k_{1}}{\underbrace{\alpha _{1}\cdots \alpha _{1}}}
\dots
\dots
\underset{%
k_{m}}{\underbrace{\alpha _{m}\cdots \alpha _{m}}}}},
\end{eqnarray}
and redefine the ${\small G'_{\dots }} \equiv \frac{{\small G}_{\dots }}{P}$ as the new 
${\small G}_{\dots }$.\footnote{For the sake of keeping the minimal amount of notations, later in all sections, 
we will map the genotype population ${\small G}_{\dots }$ to genotype frequency ${\small G'_{\dots }}$,
${\small G'_{\dots }} \equiv \frac{{\small G}_{\dots }}{P}$, then rename the genotype frequency as ${\small G_{\dots }}$.}
In the discussion below, we will  explicitly write down the coefficients for the simple case of $\N=1,2$, as in \Sec{sec:level2}, \Sec{sec:level3}  and in \cite{1410.3456}.
We can also 
explicitly write down the coefficients of the governing equations for N-gender mating  (for $\N=3$ in \Sec{sec:level4} and for a general N in \Sec{sec:level6})
from the coefficient of ${\N \choose k_0, k_1, \ldots, k_n}$ of multinomial theorem, but we leave this coefficient implicit to simplify the notations.
Another remark is that our governing equation shall also be viewed as generalizing the diploid system to the N-polypoid system.
For the N-gender-mating N-polypoid system, if the previous four assumptions are obeyed, 
we prove that the global stability of genotype frequencies can be shown
to be robust even for N-gender-mating.

%

\section{\label{sec:level2}1-gender gene-mating model (asexual)}

Here we start from the asexual mating system. There exists only one
gender in a system, such that each gender could self-produce its offspring
exactly the same as itself. Then we denote the population number of certain
genotype as ${\small G}_{\alpha }$, where $\alpha $ is
arbitrary integer number. 
Thus, parallel to the work \cite{1410.3456}, based on the assumptions done in Sec.\ref{sec:level1},
we can derive the \textbf{governing equations for population}
\begin{equation}
\frac{d}{dt}{\small G}_{\alpha }{\small =k_{b}G}_{\alpha }{\small -k}_{d}%
{\small G}_{\alpha },
\end{equation}
where ${\small k_{b}}$ is birth rate and ${\small k}_{d}$ is death rate.
Apparently we can normalize the ${\small G}_{\alpha }$ by the total population $P$,
and we will relabel 
$$\frac{{\small G}_{\alpha }}{P} \to {\small G}_{\alpha }$$ as the genotype frequency. 
Hereafter we will re-parameterize the equation in terms of the genotype frequency ${\small G}_{\alpha }$,
and we obtain the \textbf{governing equations for genotype frequencies}
\begin{equation}
\frac{d}{dt}{\small G}_{\alpha }{\small =0}.
\end{equation}
Hence, any point in the genotype frequency parameter space is stable for 1-gender asexual system.
The genotype frequency is within the range of $[0,1]$.

\section{\label{sec:level3}2-gender gene-mating model}

This section briefly summarizes the analytic solution studied 
in our previous work\cite{1410.3456}. 
We had studied
the \textbf{governing equations for genotype frequencies}:
\begin{equation}
\bigskip \left\{ 
\begin{array}{l}
\frac{d}{dt}{\small G}_{\alpha \alpha }{\small =k}_{b}{\small (G}_{\alpha
\alpha }\overset{n}{\underset{j=0}{\sum }}{\small G}_{\alpha j}{\small +}%
\frac{1}{4}\underset{i\neq \alpha }{\underset{i=0}{\overset{n}{\sum }}}%
\underset{j\neq \alpha }{\overset{n}{\underset{j=0}{\sum }}}{\small G}%
_{\alpha i}{\small G}_{\alpha j}{\small -G}_{\alpha \alpha }{\small )}, \\ 
\frac{d}{dt}{\small G}_{\alpha \beta }{\small =k}_{b}{\small (2G}_{\alpha
\alpha }{\small G}_{\beta \beta }{\small +}\underset{i\neq \alpha }{\overset{%
n}{\underset{i=0}{\sum }}}{\small G}_{\alpha i}{\small G}_{\beta \beta } 
{\small +}\underset{j\neq \beta }{%
\overset{n}{\underset{j=0}{\sum }}}{\small G}_{\alpha \alpha }{\small G}%
_{\beta j}{\small +}\frac{1}{2}\underset{i\neq \alpha }{\underset{i=0}{%
\overset{n}{\sum }}}\underset{j\neq \beta }{\overset{n}{\underset{j=0}{\sum }%
}}{\small G}_{\alpha i}{\small G}_{\beta j}{\small -G}_{\alpha \beta }%
{\small )}.%
\end{array}%
\right.
\end{equation}
Recall that the death rate $k_d$ has no net effect on the genotype frequencies, because it is a universal linear effect on each genotype population
which does not change genotype frequency.
Note that ${\small G}_{\alpha \alpha }$ means the genotype frequency contributed from the same alleles $\alpha$ from both parents.
${\small G}_{\alpha \beta }$ means the genotype frequency contributed from the alleles $\alpha$ and 
 $\beta$ from each of the parents respectively. Hereafter we will 
identify ${\small G}_{\alpha \beta }$ and its ${\small G}_{\beta \alpha}$ as the same genotype frequency,
therefore the governing equations have taken account of the combined effects of ${\small G}_{\alpha \beta }$ and its ${\small G}_{\beta \alpha}$.
We had found an exact analytic \textbf{parameterization of the global stable manifold}:
\begin{equation}
\left\{ 
\begin{array}{l}
{\small G}_{\alpha \alpha ,eq}\left( \theta _{\mu }\right) {\small =[}%
\underset{i=1}{\overset{\alpha }{\Pi }}\frac{\sin ^{2}(\theta _{i})}{(\cos
(\theta _{i})+\sin (\theta _{i}))^{2}}{\small ]}\frac{\cos ^{2}(\theta
_{\alpha +1})}{(\cos (\theta _{\alpha +1})+\sin (\theta _{\alpha +1}))^{2}},
\\ 
{\small G}_{\alpha \beta ,eq}\left( \theta _{\mu }\right) {\small =[}%
\underset{i=1}{\overset{\alpha }{\Pi }}\frac{\sin ^{2}(\theta _{i})}{(\cos
(\theta _{i})+\sin (\theta _{i}))^{2}}{\small ]}\frac{2\cos (\theta _{\alpha
+1})\sin (\theta _{\alpha +1})}{(\cos (\theta _{\alpha +1})+\sin (\theta
_{\alpha +1}))^{2}}{\small \cdot } 
{\small [}\underset{j=\alpha +2}{%
\overset{\beta }{\Pi }}\frac{\sin (\theta _{j})}{\cos (\theta _{j})+\sin
(\theta _{j})}{\small ]}\frac{\cos (\theta _{\beta +1})}{\cos (\theta
_{\beta +1})+\sin (\theta _{\beta +1})}.%
\end{array}%
\right.
\end{equation}
Her $\theta_i$ is within the range $[0, \pi/2]$.
More conveniently, we find that a \textbf{new parametrization of coordinates},
\begin{equation}
\left\{ 
\begin{array}{l}
G_{\alpha \alpha }=G_{\alpha \alpha ,eq}\left( \theta _{\mu }\right) +%
\overset{n}{\underset{i=0,i\neq \alpha }{\sum }}s_{\alpha i}, \\ 
G_{\alpha \beta }=G_{\alpha \beta ,eq}\left( \theta _{\mu }\right)
-2s_{\alpha \beta },%
\end{array}%
\right.
\end{equation}
providing an \textbf{inverse transformation} mapping between the two kinds of parametrizations:
\begin{equation}
\left\{ 
\begin{array}{l}
{\small \theta }_{k}{\small =}\tan ^{-1}{\small \{[}\underset{i=k}{\overset{n%
}{\sum }}{\small (2G}_{ii}{\small +}\underset{j\neq i}{\underset{j=0}{%
\overset{n}{\sum }}}{\small G}_{ij}{\small )]/(2G}_{k-1,k-1}{\small +}%
\underset{j\neq k-1}{\underset{j=0}{\overset{n}{\sum }}}{\small G}_{k-1,%
\text{ }j}{\small )\}}\text{ },\;\;\;\;\;\;  
{\small 1\leq k\leq n}\text{, with $n$ formulas} \\ 
\\ 
{\small s}_{ij}\text{ can be solved from the following equations:} \\ 
\text{ }\left\{ 
\begin{array}{l}
\tan ^{2}{\small (\theta }_{k}{\small )=}\frac{\underset{k\leq i\leq j\leq n}%
{\overset{n}{\sum }}G_{ij}-\underset{i=k}{\overset{n}{\sum }}\underset{j=0}{%
\overset{k-1}{\sum }}s_{ij}}{(G_{k-1,k-1}-\underset{i\neq k-1}{\underset{i=0}%
{\overset{n}{\sum }}}s_{k-1,\text{ }i})},\;\;\;\;\;\;  
{\small 1\leq k\leq n}%
\text{, with $n$ equations} \\ 
\tan {\small (\theta }_{k}{\small )=[}\underset{i=k}{\overset{n}{\sum }}%
{\small (G}_{mi}{\small +2s}_{mi}{\small )]/(G}_{m,k-1}{\small +2s}_{m,k-1}%
{\small )}, \;\;\;\;\;\; 
{\small 2\leq k\leq n}\text{, }{\small 0\leq m\leq k-2}\text{%
, }\text{ with }\binom{n}{2}\text{ equations}%
\end{array}%
\right.%
\end{array}%
\right.
\end{equation}

\bigskip Given any given initial values of ${\small G}_{ij}$ parameters as the input data and the starting point of time-evolution, we are able to
find the final equilibrium point solution ${\small G}_{ij,eq}$:
\begin{equation}
\left\{ 
\begin{array}{l}
G_{\alpha \alpha ,eq}=\frac{1}{4}(2G_{\alpha \alpha }+\overset{n}{\underset{%
i\neq \alpha }{\underset{i=0}{\sum }}}G_{\alpha i})^{2}, \\ 
G_{\alpha \beta ,eq}=\frac{1}{2}(2G_{\alpha \alpha }+\overset{n}{\underset{%
i\neq \alpha }{\underset{i=0}{\sum }}}G_{\alpha i})(2G_{\beta \beta }+%
\overset{n}{\underset{i\neq \beta }{\underset{i=0}{\sum }}}G_{\beta i}).%
\end{array}%
\right.
\end{equation}
By the above relation, we can simplify the inverse transformation, the
function ${\small s}_{\alpha \beta }({\small G}_{ij})$

\begin{equation}
\left\{ 
\begin{array}{l}
{\small \theta }_{\mu +1}{\small =}\tan ^{-1}{\small \{[}\underset{\alpha
=\mu +1}{\overset{n}{\sum }}{\small (2G}_{\alpha \alpha }{\small +}\underset{%
i\neq \alpha }{\underset{i=0}{\overset{n}{\sum }}}{\small G}_{\alpha i}%
{\small )]/(2G}_{\mu \mu }{\small +}\underset{i\neq \mu }{\underset{i=0}{%
\overset{n}{\sum }}}{\small G}_{\mu i}{\small )\}},\;\;\;\;\;\; 
{\small 0\leq \mu \leq n-1}\text{, with $n$ formulas}. \\ 
\\ 
{\small s}_{\alpha \beta }=\frac{1}{2}(G_{\alpha \beta
,eq}(G_{ij})-G_{\alpha \beta })
=\frac{1}{2}(\frac{1}{2}(2G_{\alpha \alpha }+\overset{n}{%
\underset{i\neq \alpha }{\underset{i=0}{\sum }}}G_{\alpha i})(2G_{\beta
\beta }+\overset{n}{\underset{i\neq \beta }{\underset{i=0}{\sum }}}G_{\beta
i})-G_{\alpha \beta }), \text{ with } 
\binom{n+1}{2}\text{ formulas}.%
\end{array}%
\right.
\end{equation}
The importance of the new parametrization is that the original governing equations are simplified to a set of 
\textbf{decoupled governing equations in the new coordinates}:
\begin{equation}
\left\{ 
\begin{array}{l}
\frac{ds_{\alpha \beta }}{dt}=-k_{b}s_{\alpha \beta },\text{ }0\leq \alpha
<\beta \leq n, \\ 
\frac{d\theta _{\mu }}{dt}=0,\text{ \ \ \ \ \ \ \ \ \ }1\leq \mu \leq n.%
\end{array}%
\right.
\end{equation}
Since the decoupled governing equations are the simple exponential-decay type differential equations, we obtain
the \textbf{exact solution of governing equations}:
\begin{eqnarray}
\overset{n}{\underset{0\leq i\leq j\leq n}{\sum }}{\small G}_{ij}{\small (t)}%
\widehat{{\small G}}_{ij} {\small =} 
\underset{0\leq i\leq j\leq n}{\overset{n}{\sum }}{\small G}_{ij,eq}{\small %
(\theta _{\mu }} {\small =} {\small \widetilde{\theta }_{\mu })}\widehat{%
{\small G}}_{ij}{\small +}\overset{n}{\underset{0\leq \alpha <\beta \leq n}{%
\sum }}\widetilde{s}_{\alpha \beta }{\small e}^{-k_{b}t}{\small (}\widehat{%
{\small G}}_{\alpha \alpha }{\small +}\widehat{{\small G}}_{\beta \beta }%
{\small -2}\widehat{{\small G}}_{\alpha \beta }{\small )}.
\end{eqnarray}
In the following sections, we will generalize the 2-gender gene-mating model to N-gender gene-mating model.

\section{\protect\bigskip \label{sec:level4}3-gender gene-mating model}
Hereafter we do not bother to write down the coefficients of the governing
equations, which could be easily determined by the multinomial theorem Eq.(\ref{eq:multinomial}), the basic mating rules and the Mendelian inheritance. 
The \textbf{governing equations of genotype frequencies} for 3-gender mating have the form:

\begin{equation}
\left\{ 
\begin{array}{l}
\frac{d}{dt}{\small G}_{\alpha \alpha \alpha }{\small =k}_{b}({\small M}%
_{i_{1}i_{2}i_{3};i_{4}i_{5}i_{6};i_{7}i_{8}i_{9}}^{\alpha \alpha \alpha }%
{\small G}_{i_{1}i_{2}i_{3}}{\small G}_{i_{4}i_{5}i_{6}}{\small G}%
_{i_{7}i_{8}i_{9}}{\small -G}_{\alpha \alpha \alpha }), \\ 
\frac{d}{dt}{\small G}_{\alpha \alpha \beta }{\small =k}_{b}({\small M}%
_{i_{1}i_{2}i_{3};i_{4}i_{5}i_{6};i_{7}i_{8}i_{9}}^{\alpha \alpha \beta }%
{\small G}_{i_{1}i_{2}i_{3}}{\small G}_{i_{4}i_{5}i_{6}}{\small G}%
_{i_{7}i_{8}i_{9}}{\small -G}_{\alpha \alpha \beta }), \\ 
\frac{d}{dt}{\small G}_{\alpha \beta \gamma }{\small =k}_{b}({\small M}%
_{i_{1}i_{2}i_{3};i_{4}i_{5}i_{6};i_{7}i_{8}i_{9}}^{\alpha \beta \gamma }%
{\small G}_{i_{1}i_{2}i_{3}}{\small G}_{i_{4}i_{5}i_{6}}{\small G}%
_{i_{7}i_{8}i_{9}}{\small -G}_{\alpha \beta \gamma }).\text{ \ \ \ \ \ \ \ \
\ \ \ \ \ \ \ }%
\end{array}%
\right.
\end{equation}
where $i_{1}\leq i_{2}\leq i_{3};$ $i_{4}\leq i_{5}\leq i_{6};$ $i_{7}\leq
i_{8}\leq i_{9}$.
We should remember that there will be a permutation factor for ${\small G}%
_{i_{1}i_{2}i_{3}}{\small G}_{i_{4}i_{5}i_{6}}{\small G}_{i_{7}i_{8}i_{9}}$,
if genotypes of three genders are not the same.
Again, we note that ${\small G}_{\alpha \beta \gamma}$ means the genotype frequency contributed from the alleles $\alpha$, 
$\beta$ and $\gamma$ from each of the three genders respectively. Hereafter we will 
identify ${\small G}_{\alpha \beta \gamma }$ and its index permutation ${\small G}_{\beta \alpha \gamma}$, ${\small G}_{\beta \gamma \alpha}$, etc., 
as the same genotype frequency. In our notation, we will define the genotype frequency ${\small G}_{\beta \alpha \gamma}$ 
such that $\alpha \leq \beta \leq \gamma $.

Interestingly, regardless the non-linear governing equations are generalized to the more-complicated 3-gender cubic form 
(instead of the 2-gender quadratic form in Sec.\ref{sec:level3}), 
we can still solve the governing equations and its global stability.
We find an analytic \textbf{parametrization of a continuous global stable manifold} as follows:
\begin{equation}
\left\{ 
\begin{array}{l}
{\small G}_{\alpha \alpha \alpha ,eq}\left( \theta _{\mu }\right) {\small =[}%
\underset{i=1}{\overset{\alpha }{\Pi }}\frac{\sin ^{3}(\theta _{i})}{(\cos
(\theta _{i})+\sin (\theta _{i}))^{3}}{\small ]}\frac{\cos ^{3}(\theta
_{\alpha +1})}{(\cos (\theta _{\alpha +1})+\sin (\theta _{\alpha +1}))^{3}},
\\ 
{\small G}_{\alpha \alpha \beta ,eq}\left( \theta _{\mu }\right) {\small =[}%
\underset{i=1}{\overset{\alpha }{\Pi }}\frac{\sin ^{3}(\theta _{i})}{(\cos
(\theta _{i})+\sin (\theta _{i}))^{3}}{\small ]}\frac{3\cos ^{2}(\theta
_{\alpha +1})\sin (\theta _{\alpha +1})}{(\cos (\theta _{\alpha +1})+\sin
(\theta _{\alpha +1}))^{3}}{\small \cdot } {\small [}\underset{j=\alpha +2%
}{\overset{\beta }{\Pi }}\frac{\sin (\theta _{j})}{\cos (\theta _{j})+\sin
(\theta _{j})}{\small ]}\frac{\cos (\theta _{\beta +1})}{\cos (\theta
_{\beta +1})+\sin (\theta _{\beta +1})}, \\ 
{\small G}_{\alpha \beta \beta ,eq}\left( \theta _{\mu }\right) {\small =[}%
\underset{i=1}{\overset{\alpha }{\Pi }}\frac{\sin ^{3}(\theta _{i})}{(\cos
(\theta _{i})+\sin (\theta _{i}))^{3}}{\small ]}\frac{3\cos (\theta _{\alpha
+1})\sin ^{2}(\theta _{\alpha +1})}{(\cos (\theta _{\alpha +1})+\sin (\theta
_{\alpha +1}))^{3}}{\small \cdot } {\small [}\underset{j=\alpha
+2}{\overset{\beta }{\Pi }}\frac{\sin ^{2}(\theta _{j})}{(\cos (\theta
_{j})+\sin (\theta _{j}))^{2}}{\small ]}\frac{\cos ^{2}(\theta _{\beta +1})}{%
(\cos (\theta _{\beta +1})+\sin (\theta _{\beta +1}))^{2}}, \\ 
{\small G}_{\alpha \beta \gamma ,eq}\left( \theta _{\mu }\right) {\small =[}%
\underset{i=1}{\overset{\alpha }{\Pi }}\frac{\sin ^{3}(\theta _{i})}{(\cos
(\theta _{i})+\sin (\theta _{i}))^{3}}{\small ]}\frac{3\cos (\theta _{\alpha
+1})\sin ^{2}(\theta _{\alpha +1})}{(\cos (\theta _{\alpha +1})+\sin (\theta
_{\alpha +1}))^{3}}{\small \cdot } {\small [}\underset{j=\alpha
+2}{\overset{\beta }{\Pi }}\frac{\sin ^{2}(\theta _{j})}{(\cos (\theta
_{j})+\sin (\theta _{j}))^{2}}{\small ]}\frac{2\cos (\theta _{\beta +1})\sin
(\theta _{\beta +1})}{(\cos (\theta _{\beta +1})+\sin (\theta _{\beta
+1}))^{2}}{\small \cdot } \\ 
\text{ \ \ \ \ \ \ \ \ \ \ \ \ \ \ \ \ \ \ \ }{\small [}\underset{k=\beta +2}%
{\overset{\gamma }{\Pi }}\frac{\sin (\theta _{k})}{(\cos (\theta _{k})+\sin
(\theta _{k}))}{\small ]}\frac{\cos (\theta _{\gamma +1})}{\cos (\theta
_{\gamma +1})+\sin (\theta _{\gamma +1})}.%
\end{array}%
\right.
\end{equation}
For the later convenience to decouple the governing equations, we find that a \textbf{new parametrization of coordinates}:
\begin{equation}
\left\{ 
\begin{array}{l}
G_{\alpha \alpha \alpha }=G_{\alpha \alpha \alpha ,eq}\left( \theta _{\mu
}\right) +\underset{i=0,i\neq \alpha }{\overset{n}{\sum }}s_{1\text{ }\alpha
i}+\underset{i=0,i\neq \alpha }{\overset{n}{\sum }} \text{sgn}(i-\alpha )
\text{ }u_{1\text{ }\alpha i}+\overset{n}{\underset{i,j\neq \alpha }{%
\underset{0\leq i<j\leq n}{\sum }}}m_{\alpha ij}, \\ 
G_{\alpha \alpha \beta }=G_{\alpha \alpha \beta ,eq}\left( \theta _{\mu
}\right) -s_{1\text{ }\alpha \beta }-3u_{1\text{ }\alpha \beta }, \\ 
G_{\alpha \beta \beta }=G_{\alpha \beta \beta ,eq}\left( \theta _{\mu
}\right) -s_{1\text{ }\alpha \beta }+3u_{1\text{ }\alpha \beta }, \\ 
G_{\alpha \beta \gamma }=G_{\alpha \beta \gamma ,eq}\left( \theta _{\mu
}\right) -3m_{\alpha \beta \gamma },%
\end{array}%
\right.
\end{equation}
where $\alpha \leq \beta \leq \gamma$.
Given any initial input ${\small G}_{ijk}$ parameters, we are able to find
the final equilibrium ${\small G}_{ijk,eq}$: 
\begin{equation}
\left\{ 
\begin{array}{l}
G_{\alpha \alpha \alpha ,eq}=(\frac{1}{3})^{3}(3G_{\alpha \alpha \alpha }+2%
\overset{n}{\underset{i\neq \alpha }{\underset{i=0}{\sum }}}G_{\alpha \alpha
i}+\overset{n}{\underset{i,j\neq \alpha }{\underset{0\leq i\leq j}{\sum }}}%
G_{\alpha ij})^{3} \\ 
G_{\alpha \alpha \beta ,eq}=\frac{3!}{2!1!}(\frac{1}{3})^{3}(3G_{\alpha
\alpha \alpha }+2\overset{n}{\underset{i\neq \alpha }{\underset{i=0}{\sum }}}%
G_{\alpha \alpha i}+\overset{n}{\underset{i,j\neq \alpha }{\underset{0\leq
i\leq j}{\sum }}}G_{\alpha ij})^{2}(3G_{\beta \beta \beta }+2\overset{n}{%
\underset{i\neq \alpha }{\underset{i=0}{\sum }}}G_{\beta \beta i}+\overset{n}%
{\underset{i,j\neq \alpha }{\underset{0\leq i\leq j}{\sum }}}G_{\beta ij})
\\ 
G_{\alpha \beta \gamma ,eq}=\frac{3!}{1!1!1!}(\frac{1}{3})^{3}(3G_{\alpha
\alpha \alpha }+2\overset{n}{\underset{i\neq \alpha }{\underset{i=0}{\sum }}}%
G_{\alpha \alpha i}+\overset{n}{\underset{i,j\neq \alpha }{\underset{0\leq
i\leq j}{\sum }}}G_{\alpha ij})(3G_{\beta \beta \beta }+2\overset{n}{%
\underset{i\neq \alpha }{\underset{i=0}{\sum }}}G_{\beta \beta i}+\overset{n}%
{\underset{i,j\neq \alpha }{\underset{0\leq i\leq j}{\sum }}}G_{\beta ij})
\\ 
\text{ \ \ \ \ \ \ \ \ \ \ \ \ \ \ \ }\cdot (3G_{\gamma \gamma \gamma }+2%
\overset{n}{\underset{i\neq \alpha }{\underset{i=0}{\sum }}}G_{\gamma \gamma
i}+\overset{n}{\underset{i,j\neq \alpha }{\underset{0\leq i\leq j}{\sum }}}%
G_{\gamma ij})%
\end{array}%
\right.
\end{equation}
By the above relation, we can define the \textbf{inverse transformation}
 from the old
to the new coordinates:
\begin{equation}
\left\{ 
\begin{array}{l}
{\small \theta }_{\mu +1}{\small =}\tan ^{-1}{\small \{[}\underset{\alpha
=\mu +1}{\overset{n}{\sum }}{\small (3G}_{\alpha \alpha \alpha }{\small +2%
\overset{n}{\underset{i\neq \alpha }{\underset{i=0}{\sum }}}G_{\alpha \alpha
i}+\overset{n}{\underset{i,j\neq \alpha }{\underset{0\leq i\leq j}{\sum }}}%
G_{\alpha ij})]/(3G_{\mu \mu \mu }+2\overset{n}{\underset{i\neq \mu }{%
\underset{i=0}{\sum }}}G_{\mu \mu i}+\overset{n}{\underset{i,j\neq \mu }{%
\underset{0\leq i\leq j}{\sum }}}G_{\mu ij})\}}, 
\;\;\; {\small 0\leq \mu \leq n-1}\text{%
, $n$ formulas}. \\ 
{\small s}_{1\text{ }\alpha \beta }=\frac{1}{2}(G_{\alpha \alpha \beta
,eq}(G_{ijk})+G_{\alpha \beta \beta ,eq}(G_{ijk})-G_{\alpha \alpha \beta
}-G_{\alpha \beta \beta }), \;\;\;   \text{  with } \binom{n+1}{2}\text{ formulas}. \\ 
{\small u}_{1\text{ }\alpha \beta }=\frac{1}{6}(G_{\alpha \beta \beta
}-G_{\alpha \alpha \beta }-G_{\alpha \beta \beta ,eq}(G_{ijk})+G_{\alpha
\alpha \beta ,eq}(G_{ijk})), \;\;\;  \text{  with }
\binom{n+1}{2}\text{ formulas}. \\ 
m_{\alpha \beta \gamma }=\frac{1}{3}(G_{\alpha \beta \gamma ,eq}\left(
G_{ijk}\right) -G_{\alpha \beta \gamma }), \;\;\; \text{  with } \binom{n+1}{3}\text{ formulas}.%
\end{array}%
\right.
\end{equation}
Our next key step is deriving the \textbf{decoupled governing equations in the new coordinates}:%
\begin{equation}
\left\{ 
\begin{array}{l}
\frac{ds_{1\text{ }\alpha \beta }}{dt}=-k_{b}s_{1\text{ }\alpha \beta },%
\text{  }0\leq \alpha <\beta \leq n, \\ 
\frac{du_{1\text{ }\alpha \beta }}{dt}=-k_{b}u_{1\text{ }\alpha \beta },%
\text{  }0\leq \alpha <\beta \leq n, \\ 
\frac{dm_{\alpha \beta \gamma }}{dt}=-k_{b}m_{\alpha \beta \gamma },\text{  }%
0\leq \alpha <\beta <\gamma \leq n, \\ 
\frac{d\theta _{\mu }}{dt}=0,\text{ \ \ \ \ \ \ \ \ \ \ \ \ \ \ }1\leq \mu
\leq n.%
\end{array}%
\right.
\end{equation}
Finally we obtain the \textbf{exact analytic solution of governing equations} for a 3-gender mating system:
\begin{align}
\underset{0\leq i\leq j\leq k\leq n}{\overset{n}{\sum }}{\small G}_{ijk}%
{\small (t)}\widehat{{\small G}}_{ijk} {\small =}  
\underset{0\leq i\leq j\leq k\leq n}{\overset{n}{\sum }}{\small G}_{ijk,eq}%
{\small (\theta _{\mu }} {\small =}{\small \widetilde{\theta }_{\mu })%
\widehat{G}_{ijk}+}\underset{0\leq \alpha <\beta \leq n}{\overset{n}{\sum }}%
\widetilde{s}_{1\text{ }\alpha \beta }{\small e}^{-k_{b}t}{\small (}\widehat{%
{\small G}}_{\alpha \alpha \alpha }{\small -}\widehat{{\small G}}_{\alpha
\alpha \beta }{\small -}\widehat{{\small G}}_{\alpha \beta \beta }{\small +}%
\widehat{{\small G}}_{\beta \beta \beta }{\small )}   \\
{\small +} \overset{n}{\underset{0\leq \alpha <\beta \leq n}{\sum }}%
\widetilde{u}_{1\text{ }\alpha \beta }{\small e}^{-k_{b}t}{\small (}\widehat{%
{\small G}}_{\alpha \alpha \alpha }{\small -3}\widehat{{\small G}}_{\alpha
\alpha \beta }{\small +3}\widehat{{\small G}}_{\alpha \beta \beta }{\small -}%
\widehat{{\small G}}_{\beta \beta \beta }{\small )} 
 {\small +}\text{ }\underset{0\leq \alpha <\beta <\gamma \leq n}{\overset{n}%
{\sum }}\text{\ }\widetilde{m}_{\alpha \beta \gamma }{\small e}^{-k_{b}t}(%
\widehat{{\small G}}_{\alpha \alpha \alpha }{\small +}\widehat{{\small G}}%
_{\beta \beta \beta }+\widehat{{\small G}}_{\gamma \gamma \gamma }-3\widehat{%
{\small G}}_{\alpha \beta \gamma }).  \notag
\end{align}
Due to the exponential decay to the equilibrium manifold, we have proved the global stability of 3-gender mating system.

\section{\protect\bigskip \label{sec:level6}N-gender gene-mating model}
Now we will generalize the above analysis to the N-gender gene-mating model.
Each genotype frequency parameter can be written as,
$G_{\underset{n_{1}}{\underbrace{\alpha _{1}\cdots \alpha _{1}}}\underset{%
n_{2}}{\underbrace{\alpha _{2}\cdots \alpha _{2}}}\underset{n_{3}}{%
\underbrace{\alpha _{3}\cdots \alpha _{3}}}{\small \cdots \cdots }\underset{%
n_{m}}{\underbrace{\alpha _{m}\cdots \alpha _{m}}}}$,
each is within the range $[0,1]$.  
We find the existence of the global stable fixed points as a continuous manifold and the
\textbf{parametrization of the global stable manifold} as:

\begin{equation}
\left\{ 
\begin{array}{l}
G_{\underset{n_{1}}{\underbrace{\alpha _{1}\cdots \alpha _{1}}}\underset{%
n_{2}}{\underbrace{\alpha _{2}\cdots \alpha _{2}}}{\small \cdots \cdots }%
\underset{n_{m}}{\underbrace{\alpha _{m}\cdots \alpha _{m}}}}\left( \theta
_{\mu }\right) =(\overset{\alpha _{1}}{\underset{i_{1}=1}{\prod }}\frac{\sin
^{\N}(\theta _{i_{1}})}{(\cos (\theta _{i_{1}})+\sin (\theta _{i_{1}}))^{\N}})%
\frac{\binom{\N}{n_{1}}\cos ^{n1}(\theta _{\alpha _{1}+1})\sin
^{\N-n_{1}}(\theta _{\alpha _{1}+1})}{(\cos (\theta _{\alpha _{1}+1})+\sin
(\theta _{\alpha _{1}+1}))^{\N}}{\small \cdot }\\
(\overset{\alpha _{2}}{\underset{i_{2}=\alpha _{1}+2}{\prod }}\frac{\sin
^{\N-n_{1}}(\theta _{i_{2}})}{(\cos (\theta _{i_{2}})+\sin (\theta
_{i_{2}}))^{\N-n_{1}}})\frac{\binom{\N-n_{1}}{n_{2}}\cos ^{n_{2}}(\theta
_{\alpha _{2}+1})\sin ^{\N-n_{1}-n_{2}}(\theta _{\alpha _{2}+1})}{(\cos
(\theta _{\alpha _{2}+1})+\sin (\theta _{\alpha _{2}+1}))^{\N-n_{1}}}{\small %
\cdot }\\
{\small \cdots \cdot }\ (\overset{\alpha _{m}}{\underset{i_{m}=\alpha
_{m-1}+2}{\prod }}\frac{\sin ^{n_{m}}(\theta _{i_{m}})}{(\cos (\theta
_{i_{m}})+\sin (\theta _{i_{m}}))^{n_{m}}})\frac{\cos ^{n_{m}}(\theta
_{\alpha _{m}+1})}{(\cos (\theta _{\alpha _{m}+1})+\sin (\theta _{\alpha
_{m}+1}))^{n_{m}}},\\
 {\scriptsize G}_{\underset{n_{1}}{\underbrace{\alpha _{1}\cdots
\alpha _{1}}} 
{\small \cdots \cdots }\underset{n_{m}}{\underbrace{\alpha _{m}\cdots \alpha
_{m}}}}\left( \theta _{\mu }\right) {\scriptsize =}\overset{m}{\underset{k=1}%
{\prod }}{\scriptsize [(}\overset{\alpha _{k}}{\underset{i=\alpha _{k-1}+2}{%
\prod }}{\scriptsize (}\frac{\sin (\theta _{i})}{\cos (\theta _{i})+\sin
(\theta _{i})}{\scriptsize )}^{^{\N-\overset{{\small k-1}}{\underset{{\small %
j=1}}{\Sigma }}n_{j}}}\binom{\N-\overset{{\small k-1}}{\underset{{\small j=1}}%
{\Sigma }}n_{j}}{n_{k}}\frac{\cos ^{n_{k}}(\theta _{\alpha _{k}+1})\sin ^{\N-%
\overset{{\small k}}{\underset{{\small j=1}}{\Sigma }}n_{j}}(\theta _{\alpha
_{k}+1})}{(\cos (\theta _{\alpha _{k}+1})+\sin (\theta _{\alpha
_{k}+1}))^{^{\N-\overset{{\small k-1}}{\underset{{\small j=1}}{\Sigma }}%
n_{j}}}}{\scriptsize )]}.
\end{array}%
\right.
\end{equation}
where we define $\alpha _{0}\equiv -1$.
All $\theta_\mu$ are within the range $[0,\pi/2]$.

We extend the stable manifold through the following parametrized Euclidean
fiber at each fixed point on the stable manifold.
This procedure lead to our \textbf{new parametrization of coordinates}.
For N-gender mating of $(n+1)$ alleles, we separate the parametrization of
fiber space to two cases, for N is an even integer and for N is an odd integer. 

\noindent
$\bullet$ {\bf (1) For $\N\in $ even, $\N=2k$}:\\
There are 3 types of vectors spanning the extended Euclidean bundle space.\\
(a) The symmetry type for $\alpha \leftrightarrow \beta$:
\begin{equation}
\widehat{{\small s}}_{i\text{ }\alpha \beta }=\widehat{G}_{\underset{2k+1-i}{%
\underbrace{\alpha \cdots \alpha }}\underset{i-1}{\underbrace{\beta \cdots
\beta }}}-2\widehat{G}_{\underset{k}{\underbrace{\alpha \cdots \alpha }}%
\underset{k}{\underbrace{\beta \cdots \beta }}}\bigskip +\widehat{G}_{%
\underset{i-1}{\underbrace{\alpha \cdots \alpha }}\underset{2k+1-i}{%
\underbrace{\beta \cdots \beta }}},\text{where }1\leq i\leq k.
\end{equation}
(b) Th anti-symmetry type for $\alpha \leftrightarrow \beta$:\\
\begin{equation}
\widehat{{\small u}}_{i\text{ }\alpha \beta }=\widehat{G}_{\underset{2k+1-i}{%
\underbrace{\alpha \cdots \alpha }}\underset{i-1}{\underbrace{\beta \cdots
\beta }}}-(k+1-i)\widehat{G}_{\underset{k+1}{\underbrace{\alpha \cdots
\alpha }}\underset{k-1}{\underbrace{\beta \cdots \beta }}}\bigskip +(k+1-i)%
\widehat{G}_{\underset{k-1}{\underbrace{\alpha \cdots \alpha }}\underset{k+1}%
{\underbrace{\beta \cdots \beta }}}-\widehat{G}_{\underset{i-1}{\underbrace{%
\alpha \cdots \alpha }}\underset{2k+1-i}{\underbrace{\beta \cdots \beta }}},%
\text{where }1\leq i\leq k-1.
\end{equation}
(c) The mixed of multiple-genes (mixed type number $\geq 3$) vector:\\
\begin{eqnarray}
\widehat{m}_{\underset{n_{1}}{\underbrace{\alpha _{1}\cdots \alpha _{1}}}%
\underset{n_{2}}{\underbrace{\alpha _{2}\cdots \alpha _{2}}}{\small \cdots
\cdots }\underset{n_{m}}{\underbrace{\alpha _{m}\cdots \alpha _{m}}}}
&=&n_{1}\widehat{G}_{\alpha _{1}\cdots \alpha _{1}}+n_{2}\widehat{G}_{\alpha
_{2}\cdots \alpha _{2}}+\cdots +n_{m}\widehat{G}_{\alpha _{m}\cdots \alpha
_{m}}-n\widehat{G}_{\underset{n_{1}}{\underbrace{\alpha _{1}\cdots \alpha
_{1}}}\underset{n_{2}}{\underbrace{\alpha _{2}\cdots \alpha _{2}}}{\small %
\cdots \cdots }\underset{n_{m}}{\underbrace{\alpha _{m}\cdots \alpha _{m}}}},
\notag \\
\text{ \ \ \ \ \ where }n &=&n_{1}+n_{2}+\cdots +n_{m}=\overset{m}{\underset{%
i=1}{\sum }}n_{i}.
\end{eqnarray}

\bigskip

\noindent
$\bullet$ {\bf (2) $\N\in $ odd, $\N=2k+1$}:
There are 3 types of vectors spanning the extended Euclidean bundle space.\\
(a) The symmetry type for $\alpha \leftrightarrow \beta$:\\
\begin{equation}
\widehat{{\small s}}_{i\text{ }\alpha \beta }=\widehat{G}_{\underset{2k+2-i}{%
\underbrace{\alpha \cdots \alpha }}\underset{i-1}{\underbrace{\beta \cdots
\beta }}}-\widehat{G}_{\underset{k+1}{\underbrace{\alpha \cdots \alpha }}%
\underset{k}{\underbrace{\beta \cdots \beta }}}\bigskip -\widehat{G}_{%
\underset{k}{\underbrace{\alpha \cdots \alpha }}\underset{k+1}{\underbrace{%
\beta \cdots \beta }}}+\widehat{G}_{\underset{i-1}{\underbrace{\alpha \cdots
\alpha }}\underset{2k+2-i}{\underbrace{\beta \cdots \beta }}},\text{where }%
1\leq i\leq k.
\end{equation}
(b) The anti-symmetry for $\alpha \leftrightarrow \beta$:\\
\begin{equation}
\widehat{{\small u}}_{i\text{ }\alpha \beta }=\widehat{G}_{\underset{2k+2-i}{%
\underbrace{\alpha \cdots \alpha }}\underset{i-1}{\underbrace{\beta \cdots
\beta }}}-(2k+3-2i)\widehat{G}_{\underset{k+1}{\underbrace{\alpha \cdots
\alpha }}\underset{k}{\underbrace{\beta \cdots \beta }}}\bigskip +(2k+3-2i)%
\widehat{G}_{\underset{k}{\underbrace{\alpha \cdots \alpha }}\underset{k+1}{%
\underbrace{\beta \cdots \beta }}}-\widehat{G}_{\underset{i-1}{\underbrace{%
\alpha \cdots \alpha }}\underset{2k+2-i}{\underbrace{\beta \cdots \beta }}},%
\text{where }1\leq i\leq k.
\end{equation}
(c) The vector of mixed multiple-genes(mixed type number$\geq 3$):\\
\begin{eqnarray}
\widehat{m}_{\underset{n_{1}}{\underbrace{\alpha _{1}\cdots \alpha _{1}}}%
\underset{n_{2}}{\underbrace{\alpha _{2}\cdots \alpha _{2}}}{\small \cdots
\cdots }\underset{n_{m}}{\underbrace{\alpha _{m}\cdots \alpha _{m}}}}
&=&n_{1}\widehat{G}_{\alpha _{1}\cdots \alpha _{1}}+n_{2}\widehat{G}_{\alpha
_{2}\cdots \alpha _{2}}+\cdots +n_{m}\widehat{G}_{\alpha _{m}\cdots \alpha
_{m}}-\N\widehat{G}_{\underset{n_{1}}{\underbrace{\alpha _{1}\cdots \alpha
_{1}}}\underset{n_{2}}{\underbrace{\alpha _{2}\cdots \alpha _{2}}}{\small %
\cdots \cdots }\underset{n_{m}}{\underbrace{\alpha _{m}\cdots \alpha _{m}}}},
\notag \\
\text{ \ \ where } \N &=&n_{1}+n_{2}+\cdots +n_{m}=\overset{m}{\underset{i=1}{%
\sum }}n_{i}.
\end{eqnarray}
For example, we denote $\underset{i=0}{\overset{\N}{\sum }}c(i)$ $\widehat{G}_{\underset{%
N-i}{\underbrace{\alpha \alpha }}\underset{i}{\underbrace{\beta \cdots \beta 
}}}$ as $\left( c(0),c\left( 1\right) ,\cdots ,c(n)\right)$.\\
\noindent
For $\N=2$, we have $\widehat{{\small s}}_{1\text{ }\alpha \beta }=(1,-2,1)$.\\
For $\N=3$, we have $\widehat{{\small s}}_{1\text{ }\alpha \beta }=(1,-1,-1,1)$, $\widehat{{\small u}}_{1\text{ }\alpha \beta }=(1,-3,3,-1)$.\\
For $\N=4$, we have $\widehat{{\small s}}_{2\text{ }\alpha \beta }=(0,1,-2,1,0)$, $\widehat{{\small s}}_{1\text{ }\alpha \beta }=(1,0,-2,0,1)$,
$\widehat{{\small u}}_{1\text{ }\alpha \beta }=(1,-2,0,2,-1)$.\\
For $\N=5$,  we have $\widehat{{\small s}}_{2\text{ }\alpha \beta }=(0,1,-1,-1,1,0)$,
$\widehat{{\small s}}_{1\text{ }\alpha \beta }=(1,0,-1,-1,0,1)$,
$\widehat{{\small u}}_{2\text{ }\alpha \beta }=(0,1,-3,3,-1,0)$,
$\widehat{{\small u}}_{1\text{ }\alpha \beta }=(1,0,-5,5,0,-1)$.\\
For $\N=6$, we have
$\widehat{{\small s}}_{3\text{ }\alpha \beta }=(0,0,1,-2,1,0,0)$,
$\widehat{{\small s}}_{2\text{ }\alpha \beta }=(0,1,0,-2,0,1,0)$,
$\widehat{{\small s}}_{1\text{ }\alpha \beta }=(1,0,0,-2,0,0,1)$,
$\widehat{{\small u}}_{2\text{ }\alpha \beta }=(0,1,-2,0,2,-1,0)$,
$\widehat{{\small u}}_{1\text{ }\alpha \beta }=(1,0,-3,0,3,0,-1)$.

The parametrization of the genotype frequency space for the N-gender gene-mating of (n+1) alleles system is like
the parametrization of the fiber bundle space including both the global stable base manifold and the fibers attached on the manifold.
We obtain:
\begin{eqnarray}
&&\overset{n}{\underset{0\leq i_{1}\leq \cdots \leq i_{\N}\leq n}{\sum }}%
G_{i_{1}i_{2}i_{3}\cdots i_{\N-2}i_{\N-1}i_{\N}}\widehat{G}_{i_{1}i_{2}i_{3}%
\cdots i_{\N-2}i_{\N-1}i_{\N}}  \notag \\ 
&=&
\overset{n}{\underset{0\leq i_{1}\leq \cdots \leq i_{\N}\leq n}{\sum }}%
G_{i_{1}i_{2}i_{3}\cdots i_{\N-2}i_{\N-1}i_{\N},eq}\left( \theta _{\mu }\right) 
\widehat{G}_{i_{1}i_{2}i_{3}\cdots i_{\N-2}i_{\N-1}i_{\N}} 
+\overset{[\frac{\N}{2}]}{\underset{i=1}{\sum }}\underset{0\leq \alpha \leq
\beta \leq n}{\overset{n}{\sum }}{\small s}_{i\text{ }\alpha \beta }%
\widehat{{\small s}}_{i\text{ }\alpha \beta }+\overset{[\frac{\N-1}{2}]}{%
\underset{i=1}{\sum }}\underset{0\leq \alpha \leq \beta \leq n}{\overset{n}{%
\sum }}{\small u}_{i\text{ }\alpha \beta }\widehat{{\small u}}_{i\text{ }%
\alpha \beta } +  \notag \\
&&\underset{0\leq \alpha _{1}<\cdots <\alpha _{n}\leq n}{\overset{n}{\sum }}%
m_{\underset{n_{1}}{\underbrace{\alpha _{1}\cdots \alpha _{1}}}\underset{%
n_{2}}{\underbrace{\alpha _{2}\cdots \alpha _{2}}}{\small \cdots \cdots }%
\underset{n_{m}}{\underbrace{\alpha _{m}\cdots \alpha _{m}}}}\widehat{m}_{%
\underset{n_{1}}{\underbrace{\alpha _{1}\cdots \alpha _{1}}}\underset{n_{2}}{%
\underbrace{\alpha _{2}\cdots \alpha _{2}}}{\small \cdots \cdots }\underset{%
n_{m}}{\underbrace{\alpha _{m}\cdots \alpha _{m}}}}.
\end{eqnarray}


Given any initial values of the genotype frequency ${\small G}_{\dots}$ parameters, we are able to find
the final equilibrium ${\small G}_{\dots,eq}$ via the \textbf{inverse transformation}:
\begin{eqnarray}
G_{\underset{n_{1}}{\underbrace{\alpha _{1}\cdots \alpha _{1}}}\underset{%
n_{2}}{\underbrace{\alpha _{2}\cdots \alpha _{2}}}{\small \cdots \cdots }%
\underset{n_{m}}{\underbrace{\alpha _{m}\cdots \alpha _{m}}},eq} &=&\frac{\N !%
}{n_{1}!n_{2}!\cdots n_{m}!}(\frac{1}{\N})^{\N}(\N G_{\alpha _{1}\cdots \alpha
_{1}}+(\N-1)\overset{n}{\underset{i_{1}\neq \alpha }{\underset{i_{1}=0}{\sum }%
}}G_{\alpha _{1}\cdots \alpha _{1}i_{1}}+  \notag \\
&&(\N-2)\overset{n}{\underset{i_{1},i_{2}\neq \alpha }{\underset{0\leq
i_{1}\leq i_{2}}{\sum }}}G_{\alpha _{1}\cdots \alpha _{1}i_{1}i_{2}}+\cdots +%
\overset{n}{\underset{i_{1},\cdots ,i_{\N-1}\neq \alpha }{\underset{0\leq
i_{1}\leq \cdots \leq i_{\N-1}\leq n}{\sum }}}G_{\alpha _{1}i_{1}i_{2}\cdots
i_{\N-1}})^{n_{1}}(\cdots)^{n_{2}}\cdots (\cdots)^{n_{m}}  \notag \\
\ \ &=&\frac{\N!}{n_{1}!n_{2}!\cdots n_{m}!}(\frac{1}{\N})^{\N}\overset{m}{%
\underset{j=1}{\prod }}[\N G_{\alpha _{j}\cdots \alpha _{j}}+(\N-1)\overset{n}{%
\underset{i_{1}\neq \alpha _{j}}{\underset{i_{1}=0}{\sum }}}G_{\alpha
_{j}\cdots \alpha _{j}i_{1}}  \notag \\
&&+(\N-2)\overset{n}{\underset{i_{1},i_{2}\neq \alpha _{j}}{\underset{0\leq
i_{1}\leq i_{2}}{\sum }}}G_{\alpha _{j}\cdots \alpha _{j}i_{1}i_{2}}+\cdots +%
\overset{n}{\underset{i_{1},\cdots ,i_{\N-1}\neq \alpha _{j}}{\underset{0\leq
i_{1}\leq \cdots \leq i_{\N-1}\leq n}{\sum }}}G_{\alpha _{j}i_{1}i_{2}\cdots
i_{\N-1}}]^{n_{j}}  \notag \\
\ &=&\frac{\N!}{n_{1}!n_{2}!\cdots n_{m}!}(\frac{1}{\N})^{\N}\overset{m}{%
\underset{j=1}{\prod }}\{\underset{k=0}{\overset{\N-1}{\sum }}(\N-k)[%
\underset{i_{1},\cdots ,i_{k}\neq \alpha _{j}}{\underset{0\leq i_{1}\leq
\cdots \leq i_{k}\leq n}{\overset{n}{\sum }}}G_{\alpha _{j}\cdots \alpha
_{j}i_{1}\cdots i_{k}}]\}^{n_{j}}
\end{eqnarray}
By the above relation, we can simplify the inverse transformation from the old
to the new coordinates:
\begin{equation}
\left\{ 
\begin{array}{l}
{\small \theta }_{\mu +1}{\small =}\tan ^{-1}{\small \{[}\underset{\alpha
=\mu +1}{\overset{n}{\sum }}{\small (\N G_{\alpha \cdots \alpha }+(\N -1)\overset%
{n}{\underset{i_{1}\neq \alpha }{\underset{i_{1}=0}{\sum }}}G_{\alpha \cdots
\alpha i_{1}}+(\N -2)\overset{n}{\underset{i_{1},i_{2}\neq \alpha }{\underset{%
0\leq i_{1}\leq i_{2}}{\sum }}}G_{\alpha \cdots \alpha i_{1}i_{2}}+\cdots +%
\overset{n}{\underset{i_{1},\cdots ,i_{\N-1}\neq \alpha }{\underset{0\leq
i_{1}\leq \cdots \leq i_{\N-1}\leq n}{\sum }}}G_{\alpha i_{1}i_{2}\cdots
i_{\N-1}})]/} \\ 
\text{ \ \ \ \ \ \ \ \ \ \ \ \ \ \ \ \ \ \ \ \ \ \ \ \ \ \ \ \ \ \ \ \ \ \ }%
{\small [\N G_{\mu \cdots \mu }+(\N-1)\overset{n}{\underset{i_{1}\neq \mu }{%
\underset{i_{1}=0}{\sum }}}G_{\mu \cdots \mu i_{1}}+(\N-2)\overset{n}{%
\underset{i_{1},i_{2}\neq \mu }{\underset{0\leq i_{1}\leq i_{2}}{\sum }}}%
G_{\mu \cdots \mu i_{1}i_{2}}+\cdots +\overset{n}{\underset{i_{1},\cdots
,i_{\N-1}\neq \mu }{\underset{0\leq i_{1}\leq \cdots \leq i_{\N-1}\leq n}{\sum 
}}}G_{\mu i_{1}i_{2}\cdots i_{\N-1}}]\}} \\ 
\text{{\small For} }{\small \N=2k,}\left\{ 
\begin{array}{l}
{\small s}_{i\text{ }\alpha \beta }{\small =}\frac{1}{2}{\small (G}_{%
\underset{2k+1-i}{\underbrace{\alpha \cdots \alpha }}\underset{i-1}{%
\underbrace{\beta \cdots \beta }}}{\small +G}_{\underset{i-1}{\underbrace{%
\alpha \cdots \alpha }}\underset{2k+1-i}{\underbrace{\beta \cdots \beta }}}%
{\small -G}_{\underset{2k+1-i}{\underbrace{\alpha \cdots \alpha }}\underset{%
i-1}{\underbrace{\beta \cdots \beta }},eq}{\small -G}_{\underset{i-1}{%
\underbrace{\alpha \cdots \alpha }}\underset{2k+1-i}{\underbrace{\beta
\cdots \beta }},eq}{\small ), \;\; 2\leq i\leq k} \\ 
{\small s}_{1\text{ }\alpha \beta }{\small =}\frac{1}{2}{\small (G}_{%
\underset{k}{\underbrace{\alpha \cdots \alpha }}\underset{k}{\underbrace{%
\beta \cdots \beta }}eq}{\small -G}_{\underset{k}{\underbrace{\alpha \cdots
\alpha }}\underset{k}{\underbrace{\beta \cdots \beta }}}{\small -2}\underset{%
i=2}{\overset{k}{\sum }}{\small s}_{i\text{ }\alpha \beta }{\small )}\text{ 
} \\ 
{\small u}_{i\text{ }\alpha \beta }{\small =}\frac{1}{2}{\small (G}_{%
\underset{2k+1-i}{\underbrace{\alpha \cdots \alpha }}\underset{i-1}{%
\underbrace{\beta \cdots \beta }}}{\small -G}_{\underset{i-1}{\underbrace{%
\alpha \cdots \alpha }}\underset{2k+1-i}{\underbrace{\beta \cdots \beta }}}%
{\small +G}_{\underset{2k+1-i}{\underbrace{\alpha \cdots \alpha }}\underset{%
i-1}{\underbrace{\beta \cdots \beta }},eq}{\small -G}_{\underset{i-1}{%
\underbrace{\alpha \cdots \alpha }}\underset{2k+1-i}{\underbrace{\beta
\cdots \beta }},eq}{\small ), \;\;2\leq i\leq k-1} \\ 
{\small u}_{1\text{ }\alpha \beta }{\small =}\frac{1}{\N}{\small (G}_{%
\underset{k+1}{\underbrace{\alpha \cdots \alpha }}\underset{k-1}{\underbrace{%
\beta \cdots \beta }}eq}{\small -G}_{\underset{k-1}{\underbrace{\alpha
\cdots \alpha }}\underset{k+1}{\underbrace{\beta \cdots \beta }}eq}{\small -G%
}_{\underset{k+1}{\underbrace{\alpha \cdots \alpha }}\underset{k-1}{%
\underbrace{\beta \cdots \beta }}}{\small +G}_{\underset{k-1}{\underbrace{%
\alpha \cdots \alpha }}\underset{k+1}{\underbrace{\beta \cdots \beta }}}%
{\small -2}\overset{k-1}{\underset{i=2}{\sum }}{\small (k+1-i)u}_{i\text{ }%
\alpha \beta }{\small )}%
\end{array}%
\right. \\ 
\text{{\small For} }{\small \N=2k+1},\left\{ 
\begin{array}{l}
{\small s}_{i\text{ }\alpha \beta }{\small =}\frac{1}{2}{\small (G}_{%
\underset{2k+2-i}{\underbrace{\alpha \cdots \alpha }}\underset{i-1}{%
\underbrace{\beta \cdots \beta }}}{\small +G}_{\underset{i-1}{\underbrace{%
\alpha \cdots \alpha }}\underset{2k+2-i}{\underbrace{\beta \cdots \beta }}}%
{\small -G}_{\underset{2k+2-i}{\underbrace{\alpha \cdots \alpha }}\underset{%
i-1}{\underbrace{\beta \cdots \beta }},eq}{\small -G}_{\underset{i-1}{%
\underbrace{\alpha \cdots \alpha }}\underset{2k+2-i}{\underbrace{\beta
\cdots \beta }},eq}{\small ), \;\; 2\leq i\leq k} \\ 
{\small s}_{1\text{ }\alpha \beta }{\small =}\frac{1}{2}{\small (G}_{%
\underset{k+1}{\underbrace{\alpha \cdots \alpha }}\underset{k}{\underbrace{%
\beta \cdots \beta }}eq}{\small +G}_{\underset{k}{\underbrace{\alpha \cdots
\alpha }}\underset{k+1}{\underbrace{\beta \cdots \beta }}eq}{\small -G}_{%
\underset{k+1}{\underbrace{\alpha \cdots \alpha }}\underset{k}{\underbrace{%
\beta \cdots \beta }}}{\small -G}_{\underset{k}{\underbrace{\alpha \cdots
\alpha }}\underset{k+1}{\underbrace{\beta \cdots \beta }}}{\small -2}%
\underset{i=2}{\overset{k}{\sum }}{\small s}_{i\text{ }\alpha \beta }%
{\small )}\text{ } \\ 
{\small u}_{i\text{ }\alpha \beta }{\small =}\frac{1}{2}{\small (G}_{%
\underset{2k+2-i}{\underbrace{\alpha \cdots \alpha }}\underset{i-1}{%
\underbrace{\beta \cdots \beta }}}{\small -G}_{\underset{i-1}{\underbrace{%
\alpha \cdots \alpha }}\underset{2k+2-i}{\underbrace{\beta \cdots \beta }}}%
{\small -G}_{\underset{2k+2-i}{\underbrace{\alpha \cdots \alpha }}\underset{%
i-1}{\underbrace{\beta \cdots \beta }},eq}{\small +G}_{\underset{i-1}{%
\underbrace{\alpha \cdots \alpha }}\underset{2k+2-i}{\underbrace{\beta
\cdots \beta }},eq}{\small ), \;\; 2\leq i\leq k} \\ 
{\small u}_{1\text{ }\alpha \beta }{\small =}\frac{1}{\N}{\small (G}_{%
\underset{k+1}{\underbrace{\alpha \cdots \alpha }}\underset{k}{\underbrace{%
\beta \cdots \beta }}eq}{\small -G}_{\underset{k}{\underbrace{\alpha \cdots
\alpha }}\underset{k+1}{\underbrace{\beta \cdots \beta }}eq}{\small -G}_{%
\underset{k+1}{\underbrace{\alpha \cdots \alpha }}\underset{k}{\underbrace{%
\beta \cdots \beta }}}{\small +G}_{\underset{k}{\underbrace{\alpha \cdots
\alpha }}\underset{k+1}{\underbrace{\beta \cdots \beta }}}{\small -2}\overset%
{k}{\underset{i=2}{\sum }}{\small (2k+3-2i)u}_{i\text{ }\alpha \beta }%
{\small )}%
\end{array}%
\right. \\ 
m=\frac{1}{\N}(m_{eq}-m),\text{ where }m=m_{_{\underset{n_{1}}{\underbrace{%
\alpha _{1}\cdots \alpha _{1}}}\underset{n_{2}}{\underbrace{\alpha
_{2}\cdots \alpha _{2}}}{\small \cdots \cdots }\underset{n_{m}}{\underbrace{%
\alpha _{m}\cdots \alpha _{m}}}}}%
\end{array}%
\right.
\end{equation}
%
Next we derive the \textbf{decoupled governing equations in the new coordinates}:
\begin{equation}
\left\{ 
\begin{array}{l}
\frac{ds_{i\text{ }\alpha \beta }}{dt}=-k_{b}s_{i\text{ }\alpha \beta },%
\text{ }0\leq \alpha <\beta \leq n \\ 
\frac{du_{i\text{ }\alpha \beta }}{dt}=-k_{b}u_{i\text{ }\alpha \beta },%
\text{ }0\leq \alpha <\beta \leq n \\ 
\frac{dm}{dt}=-k_{b}m,\text{ where }m=m_{_{\underset{n_{1}}{\underbrace{%
\alpha _{1}\cdots \alpha _{1}}}\underset{n_{2}}{\underbrace{\alpha
_{2}\cdots \alpha _{2}}}{\small \cdots \cdots }\underset{n_{m}}{\underbrace{%
\alpha _{m}\cdots \alpha _{m}}}}} \\ 
\frac{d\theta _{k}}{dt}=0,\text{ \ \ \ \ \ \ \ \ \ \ \ \ \ \ }1\leq k\leq n.%
\end{array}%
\right.
\end{equation}
Parallel to Secs.\ref{sec:level3} and \ref{sec:level4}'s analysis, the decoupled governing equations are the exponential-decay type differential equations, we obtain
the \textbf{exact analytic time-evolution solution of governing equations}:
\begin{eqnarray}
&&\overset{n}{\underset{0\leq i_{1}\leq \cdots \leq i_{\N}\leq n}{\sum }}%
G_{i_{1}i_{2}i_{3}\cdots i_{\N-2}i_{\N-1}i_{\N}}(t)\widehat{G}%
_{i_{1}i_{2}i_{3}\cdots i_{\N-2}i_{\N-1}i_{\N}}   \\
&=&\overset{n}{\underset{0\leq i_{1}\leq \cdots \leq i_{\N}\leq n}{\sum }}%
G_{i_{1}i_{2}i_{3}\cdots i_{\N-2}i_{\N-1}i_{\N},eq}\left( {\small \widetilde{%
\theta }_{\mu }}\right) \widehat{G}_{i_{1}i_{2}i_{3}\cdots
i_{\N-2}i_{\N-1}i_{\N}}+  
\overset{[\frac{\N}{2}]}{\underset{i=1\quad }{\sum }}\underset{0\leq \alpha \leq
\beta \leq n}{\overset{n}{\sum }}\widetilde{{\small s}}_{i\text{ }\alpha
\beta }{\small e}^{-k_{b}t}\widehat{{\small s}}_{i\text{ }\alpha \beta } \notag \\
&&+
\overset{[\frac{\N-1}{2}]}{\underset{i=1}{\sum }}\underset{0\leq \alpha \leq
\beta \leq n}{\overset{n}{\sum }}\widetilde{{\small u}}_{i\text{ }\alpha
\beta }{\small e}^{-k_{b}t}\widehat{{\small u}}_{i\text{ }\alpha \beta }+ 
\underset{0\leq \alpha _{1}<\cdots <\alpha _{n}\leq n}{\overset{n}{\sum }}%
\widetilde{m}_{\underset{n_{1}}{\underbrace{\alpha _{1}\cdots \alpha _{1}}}%
\underset{n_{2}}{\underbrace{\alpha _{2}\cdots \alpha _{2}}}{\small \cdots
\cdots }\underset{n_{m}}{\underbrace{\alpha _{m}\cdots \alpha _{m}}}}{\small %
e}^{-k_{b}t}\widehat{m}_{\underset{n_{1}}{\underbrace{\alpha _{1}\cdots
\alpha _{1}}}\underset{n_{2}}{\underbrace{\alpha _{2}\cdots \alpha _{2}}}%
{\small \cdots \cdots }\underset{n_{m}}{\underbrace{\alpha _{m}\cdots \alpha
_{m}}}}. \notag
\end{eqnarray}

%

\section{Conclusions: Global stability of N-gender-mating polyploid systems}

\noindent
\underline{\bf 
The new findings in this work:}
We have found the exact analytic solutions and proved the global stability of
an N-gender gene-mating N-polypoid system.
Since there is no chaotic behavior, it seems to suggest that the natural law does not go against N-gender N-polypoid system as far as the global stability is concerned.
Presumably, the extraterrestrial aliens or intelligence 
may have N-gender N-polypoid system with N$\geq 3$ while the mating system still enjoys the global stability.
It is likely that when the mutation and the natural selection process sets in, it will alter the governing equations to be less-symmetric, and 
thus possibly result in richer or more chaotic behaviors. 
It will be interesting to study the perturbation away from the symmetric governing equations, with the help of our exact analytic solutions.\\

\noindent
\underline{\bf 
The new derivations and analytic solutions in this work:}
To the best of our knowledge of literature, the analysis closest to ours in the literature is Ref.\cite{Nagylaki}. 
In Ref.\cite{Nagylaki}, some exactly solvable models within a gene pool are analyzed, where the monotonic evolutionary behavior is found.
 Ref.\cite{Nagylaki}'s studies a single locus for an arbitrary number of alleles with or without distinguishing the sex.
We study a single locus, arbitrary number alleles in an N-gender multiploid (or polyploid) population.
Therefore, we generalize two degrees of freedom, N-gender and N-polyploid, and $(n+1)$-alleles, both to arbitrary integers, N and $n$.

Related models for the $2$-gender diploid can be found in \cite{book2} and a systematic derivation in Ref.~\cite{1410.3456}.
It is worthwhile mentioning that some of the previous models mostly focus on solving the ``discretized difference equations;'' 
while we focus on solving the idealized  ``highly-nonlinear coupled continuous differential equations.'' 

We can also compare the 2-gender to N-gender gene-mating equation to the Boltzmann-like equation.
The 2-gender to N-gender gene-mating equations are basically the 2-body collision to the N-body collision Boltzmann equations 
with continuous distribution functions of \emph{discretized} variables\footnote{Here the genotype label
${
  \underset{k_{0}}{\underbrace{\alpha _{0}\cdots \alpha _{0}}} %
 \underset{k_{1}}{\underbrace{\alpha _{1}\cdots \alpha _{1}}}
\dots
\dots
\underset{%
k_{m}}{\underbrace{\alpha _{m}\cdots \alpha _{m}}}}$ in
the continuous genotype frequency distribution function
${\small G_{
  \underset{k_{0}}{\underbrace{\alpha _{0}\cdots \alpha _{0}}} %
 \underset{k_{1}}{\underbrace{\alpha _{1}\cdots \alpha _{1}}}
\dots
\dots
\underset{%
k_{m}}{\underbrace{\alpha _{m}\cdots \alpha _{m}}}}}(t)$ 
is a discretized labeling, while the time $t$ in our model is continuous.
} 
instead of \emph{continuous} 
variables.\footnote{In contrast,
the conventional Boltzmann equation has the continuous variables $(\vec{x}, \vec{p})$ in the
continuous distribution function $f(\vec{x}, \vec{p},t)$,
e.g., $(\vec{x} \in \mathbb{R}, \vec{p} \in \mathbb{R})$.
}
It may be interesting to find a set of exactly solvable time-dependent Boltzmann equation by generalizing our discretized variables to continuous functional variables. Such a derived Boltzmann-like equation should still provide the global stability and
a continuous fixed-point global stable manifold in the infinite-dimensional continuous parameter space \cite{gene3}.
All these will be left open for future directions.

For more future directions, it may be enlightening to generalize our
explicit solutions for the N-gender mating without natural selection rules,
to the N-gender mating with selection rules or other modifications: 
Hofbauer and Sigmund (1988) \cite{HofbauerandSigmund1988}, 
 Akin and Szucs (1994) \cite{AkinandSzucs1994},  
Nagylaki and Crow (1974) \cite{Nagylaki} or Jost and Pepper (2008) \cite{JostPepper2008}.
%


\section{Acknowledgments}

JW  acknowledges the NSF Grant PHY-1606531 and the support from Institute for Advanced Study. 
This work is also supported by NSF Grant DMS-1607871 ``Analysis, Geometry and Mathematical
Physics'' and Center for Mathematical Sciences and Applications at Harvard University.
This work is supported by NSF
Grant No.  DMR-1005541 and NSFC 11274192.  It is also supported by the BMO
Financial Group and the John Templeton Foundation.  Research at Perimeter
Institute is supported by the Government of Canada through Industry Canada and
by the Province of Ontario through the Ministry of Research.

\end{document}